%% file: root.tex
\documentclass[conference,10pt]{IEEEtran}
 \IEEEoverridecommandlockouts
\usepackage{cite}
\usepackage{color}
\ifCLASSINFOpdf
	\usepackage[pdftex]{graphicx}
	\graphicspath{{./figures/}}
\else
	\usepackage[dvips]{graphicx}
	\graphicspath{./figures/}
\fi
\graphicspath{{fig/}{jpeg/}}
\usepackage[cmex10]{amsmath}
\usepackage{amssymb}
\usepackage{algorithm}
\usepackage{algorithmic}
\usepackage{subcaption}
\usepackage{caption}
\usepackage[outdir=./]{epstopdf}
\usepackage{comment}
\usepackage{bm}
\input{mysymbol.sty}
\input{my_sections.tex}

\usepackage{tikz}
\usetikzlibrary{shapes,arrows}
\usepackage{float,environ}
\usetikzlibrary{matrix} % for block alignment
\usetikzlibrary{decorations.markings} % for arrow heads
\usetikzlibrary{calc} % for manimulation of coordinates
\usetikzlibrary{decorations.pathmorphing,snakes, positioning, decorations.pathreplacing}

\usetikzlibrary{fit}

\usepackage{pgfplots}
\usepackage{color}
\usepackage{multirow}

\makeatletter
\newsavebox{\measure@tikzpicture}
\NewEnviron{scaletikzpicturetowidth}[1]{%
  \def\tikz@width{#1}%
  \begin{lrbox}{\measure@tikzpicture}%
  \BODY
  \end{lrbox}%
  \pgfmathparse{#1/\wd\measure@tikzpicture}%
  \BODY
}
\makeatother

\newcommand{\QED}{\hfill\ensuremath{\blacksquare}}

\def\vec2{\text{vec}}

\newtheorem{theorem}{\hspace{0pt}\bf Theorem}

\begin{document}
%
% Single address.
% ---------------
\title{Decentralized Channel Management in WLANs \\ with Graph Neural Networks %\thanks{}
}

\author{\IEEEauthorblockN{Zhan Gao$^{*}$ \qquad Yulin Shao$^{\dagger}$ \qquad Deniz G{\"u}nd{\"u}z$^{\dagger}$ \qquad Amanda Prorok$^{*}$ }
\IEEEauthorblockA{$^{*}$Department of Computer Science and Technology, University of Cambridge, Cambridge, UK \\
$^{\dagger}$Department of Electrical and Electronic Engineering, Imperial College London, London, UK  \\
Emails: \{zg292, asp45\}@cam.ac.uk, \{y.shao, d.gunduz\}@imperial.ac.uk}
}

\maketitle

\begin{abstract}
Wireless local area networks (WLANs) manage multiple access points (APs) and assign scarce radio frequency resources to APs for satisfying traffic demands of associated user devices. This paper considers the channel allocation problem in WLANs that minimizes the mutual interference among APs, and puts forth a learning-based solution that can be implemented in a decentralized manner. We formulate the channel allocation problem as an unsupervised learning problem, parameterize the control policy of radio channels with graph neural networks (GNNs), and train GNNs with the policy gradient method in a model-free manner. The proposed approach allows for a decentralized implementation due to the distributed nature of GNNs and is equivariant to network permutations. The former provides an efficient and scalable solution for large network scenarios, and the latter renders our algorithm independent of the AP reordering. Empirical results are presented to evaluate the proposed approach and corroborate theoretical findings.
\end{abstract}


\section{Introduction}

Wireless local area networks (WLANs), commonly known as Wi-Fi, are playing an increasingly important role in the wireless networking industry \cite{bellalta2016next}. According to Cisco \cite{cisco}, by 2023, the number of networked devices will be 5 billion in North America, and the number of Internet-of-things (IoT) connections is expected to reach 14.7 billion globally. In particular, $72\%$ of IoT connections will be made via Wi-Fi. The exploding number of wireless connections and advanced multimedia applications has led to an unprecedented amount of traffic over Wi-Fi networks. On the other hand, the channel resources assigned to Wi-Fi have been limited to the 2.4 GHz and 5 GHz unlicensed spectrum \cite{IEEEac}, reinforcing the need for optimizing channel allocation and network management.

Given a network of Wi-Fi access points (APs) and a fixed number of channels, channel allocation exploits spatial multiplexing and assigns the channels to the APs such that mutual interference among APs is minimized. Intuitively, a judicious channel allocation scheme assigns different channels to adjacent APs and the same channels to nonadjacent APs, thereby boosting network throughput while minimizing interference. However, the latter is not feasible for large-scale networks with a limited number of channels.

Prior solutions on WLAN channel allocation rely on a central controller to manage APs and make allocation decisions with the global network information \cite{localsearch,ILP,kaushik2021deep,TurboCA,HuaweiFrance,TokyoGCN,MADRLHT}. The problem can be formulated as a weighted graph coloring problem, which is NP-complete. To tackle the NP-hardness, various schemes, such as local-search \cite{localsearch}, integer linear programming \cite{ILP}, deep reinforcement learning (DRL) \cite{HuaweiFrance,TokyoGCN}, etc., have been proposed. Among them, three notable related works are TurboCA \cite{TurboCA}, Net2Seq \cite{HuaweiFrance}, and graph convolutional network (GCN)-based channel allocation \cite{TokyoGCN}. 

TurboCA, the commercial solution implemented in the Cisco Meraki series, is a heuristic algorithm optimized on the data collected from millions of real-world APs. The design principle of TurboCA is client-centric: APs with higher client density and usage will be assigned more channels; APs with fewer/zero clients are more inclined to switch to new channels than those with more clients; single AP failure is avoided by a multiplicative network metric design and neighborhood clearance. 
Motivated by TurboCA, the authors of \cite{HuaweiFrance} proposed Net2Seq, a DRL-based channel allocation scheme, aiming to minimize mutual interference among APs. Net2Seq tailors a few feature-extraction schemes and utilizes deep neural networks (DNNs) for decision-making. With the data-driven nature, Net2Seq is able to adapt its solution to traffic demands, thereby outperforming TurboCA in terms of minimizing network interference. In contrast, \cite{TokyoGCN} proposed GCNs to extract features directly from the interference graph of APs. In their approach, the channel allocation state and the topological information are fed into a centralized GCN, which, in turn, outputs the state-action values for all possible actions, determining the channel allocation plan \cite{DQN}. Their solution, however, is limited to small-scale problems, where each AP can only occupy a single channel, the total number of APs is few (e.g., $10$), and the interference graph is sparse.

In this paper, we put forth a decentralized learning-based solution for channel allocation in WLANs.
Our approach leverages graph neural networks (GNNs) to exploit topological insight and extract useful features from the traffic loads and channel occupations of APs (Sec. \ref{sec_gnn}).
Unlike the centralized DNN in \cite{HuaweiFrance} and GCN in \cite{TokyoGCN}, we design the GNN in a way that APs are allowed to actively select channels based on the locally sensed states of their neighbors.
Specifically, APs do not need to update the central controller about their states and passively wait for the allocation decisions; instead, they make local decisions about %autonomously choose 
which channels to occupy, yielding an efficient, scalable, and robust channel allocation solution (Sec. \ref{sec:decentralizedImplementation}).
The GNN is trained with the policy gradient method in a model-free manner, which needs only system observations, not system models; and thus, can be used in any application scenarios (Sec. \ref{sec:policyGradient}). Benefiting from the permutation equivariance of GNNs, the proposed approach is robust to AP reordering, and hence, can be implemented directly on networks with arbitrarily ordered APs (Sec. \ref{sec:permutationEquivariance}). Experiment results corroborate our theoretical findings and verify that the performance of the proposed method approaches that of the centralized solution (Sec. \ref{sec_numerical_results}).

%%%%%%%%%%%%%%%%%%%%%%%%%%%%%%%%%%%%
%%%%%%%%%%%%%%%%%%%%%%%%%%%%%%%%%%%%
%%%%  S  E  C  T  I  O  N     %%%%%%%%%%%%%%%%%%
%%%%%%%%%%%%%%%%%%%%%%%%%%%%%%%%%%%%
%%%%%%%%%%%%%%%%%%%%%%%%%%%%%%%%%%%%

\section{Problem Formulation}\label{sec:problemFormulation}

\begin{figure}
	\centering
	\includegraphics[width=0.48\textwidth, trim=0 0 0 0]{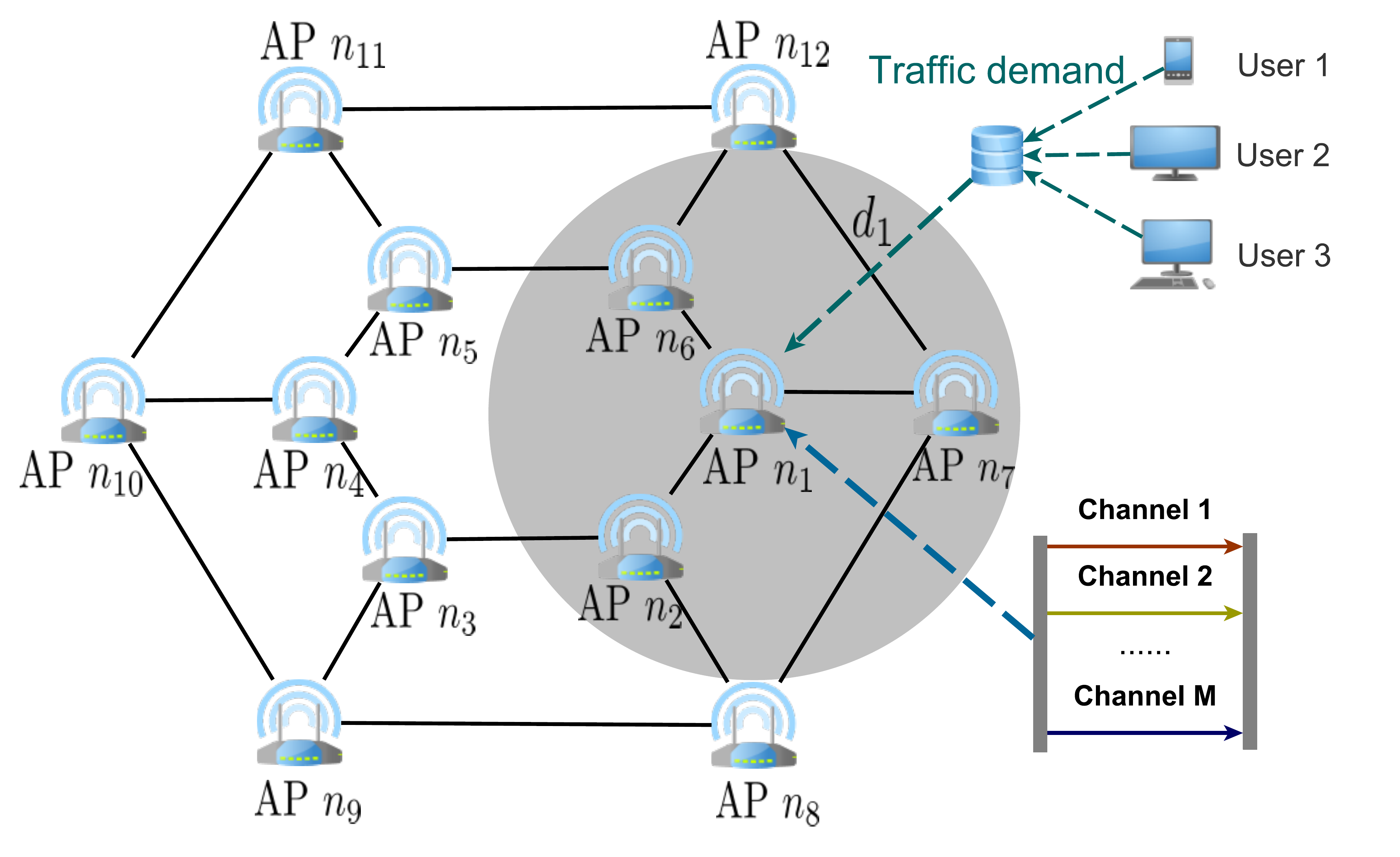}
	\caption{A WLAN with $N=12$ APs. APs are loaded with respective traffic demands of their local user devices and use different channels for data transmission. Each AP is interfered by the neighboring APs within the communication/interference range (grey circle for AP $n_1$).}
	\label{fig:WLANs}\vspace{-6mm}
\end{figure}

Consider a WLAN with $N$ APs denoted by $\{n_1,\ldots,n_N\}$ (see Fig. \ref{fig:WLANs}). The APs have different traffic demands and can choose from available radio resources, i.e., channels, to satisfy these demands. There exists mutual interference between two APs if they are audible to each other and using the same channel. %Fig. \ref{fig:WLANs} shows the details of the WLAN. 
By adaptively allocating channels among APs, our goal is to minimize the mutual interference in the system while satisfying the traffic demands of the APs.

Suppose there are %$N$ APs $\{n_1,\ldots,n_N\}$ and 
$M$ channels %$\{m_1,\ldots,m_M\}$ 
available in the network, and the traffic demands of APs are represented by $\bbd = [d_1,\ldots,d_N]^\top$ with $d_i$ denoting the demand of AP $n_i$, generated by local user devices they serve. Channels are selected by the APs based on the traffic demands $\bbd$ via a control policy $\pi(\bbd) = \bbC = [\bbc_1,\ldots,\bbc_N]$, where $\bbc_i = [c_{i1},\ldots,c_{iM}]^\top$ is the channel selection of AP $n_i$ with $c_{i\ell} \in \{0,1\}$ set to $1$ if channel $\ell$ is selected by AP $n_i$, and $0$ otherwise, for $i=1,\ldots,N$ and $\ell=1,\ldots,M$. Let $M_i$ denote the number of channels selected by AP $n_i$, i.e., $M_i = \sum_{\ell=1}^M c_{i\ell}$. 
Once the channels are selected, the traffic load of an AP is equally split on the selected channels. For example, if AP $n_i$ is loaded with traffic demand $d_i$, %and selects $1 \le M_i \le M$ channels for transmission, 
the traffic on each channel is $d_i / M_i$. 

The interference pattern among APs can be captured by a neighborhood map $\ccalN: (i,j) \to \{0,1\}$, $i,j=1,\ldots,N$, where $\ccalN(i,j) = 1$ means that the APs $n_i$ and $n_j$ are within each other's communication range.\footnote{This is commonly determined by measuring the inter-AP received signal strength and considering a threshold to establish neighborhood (e.g. $-82$ dBm) in real-world applications.} AP $n_i$ is interfered by AP $n_j$ if and only if: (i) they are within each other's communication range, i.e., $\ccalN(i,j)=1$, (ii) AP $n_j$ is transmitting, and (iii) their selected channels overlap with each other. %We remark that m
Multiple channels can be selected by a single AP to alleviate the traffic burden on each channel, although this may increase the interference between APs. This formulation implies a trade-off between the transmission burden and the interference effect.

\begin{figure*}
	\centering
	\includegraphics[width=0.8\textwidth, trim=0 0 0 0]{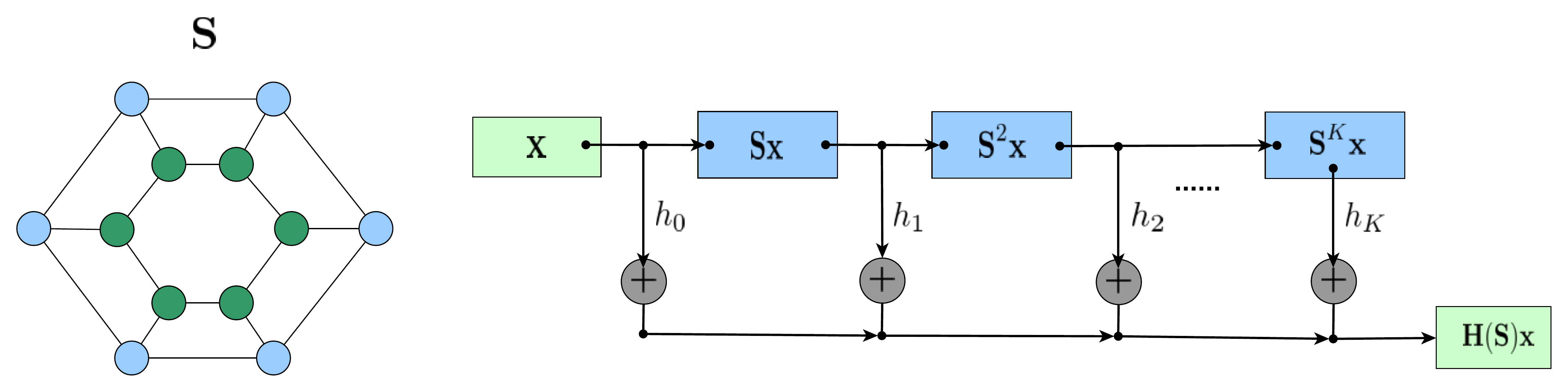}
	\caption{The GCF of order $K$. The WLAN in Fig. \ref{fig:WLANs} is modeled as a graph with a shift operator $\bbS$ (left). The GCF shifts a signal $\bbx$ over $\bbS$ and aggregates the shifted signals $\{\bbS^k\bbx\}_{k=0}^K$ with coefficients $\{h_k\}_{k=0}^K$ to generate the output $\bbH(\bbS)\bbx$ (right).}
	\label{fig:graphFilter}\vspace{-4mm}
\end{figure*}

With the above preliminaries, the interference of AP $n_j$ to $n_i$ on channel $\ell$ can be represented by a function of the traffic demands $d_i, d_j$, the neighborhood map $\ccalN(i,j)$, the channel selections $c_{i\ell}, c_{j\ell}$, and the total numbers of selected channels $M_i, M_j$ by $n_i, n_j$, i.e.,
\begin{equation}\label{eq:interferenceFunction}
	I_{i,j,\ell} \big(d_i, d_j, \ccalN(i,j), c_{i\ell}, c_{j\ell}, M_i, M_j\big).
\end{equation}
The explicit expression of function \eqref{eq:interferenceFunction} depends on specific systems and environments. Our objective is to reduce the mutual interference and balance the channel utilization over the whole network. In this context, we follow \cite{HuaweiFrance} and define the \emph{maximal channel utilization} as the interference of AP $n_i$ on the worst channel \vspace{-2mm} 
\begin{align}\label{eq:channelUtilization}
	&u_i(\bbd, \ccalN, \bbC) \\
	&= \max_{\ell = 1,\ldots,M} \sum_{j=1}^N I_{i,j,\ell}\big(d_i, d_j, \ccalN(i,j), c_{i\ell}, c_{j\ell}, M_i, M_j\big) \nonumber
\end{align} 
with $I_{i,i,\ell} = 0$ %for $i=1,\ldots,N$ 
by default. Since the traffic demands are instantaneous at APs, the mutual interference and the channel utilization are instantaneous as well. A long-term expectation is a more meaningful metric. We, thus, consider the objective function as the expected mean of $\{u_i\big(\bbd, \ccalN, \bbC\big)\}_{i=1}^N$ over APs \vspace{-1mm}
\begin{equation}\label{eq:objectiveFunction}
	\mathbb{E}_\bbd\Big[\frac{1}{N}\sum_{i=1}^N d_i u_i\big(\bbd, \ccalN, \bbC\big) \Big],
\end{equation}
where $u_i\big(\bbd, \ccalN, \bbC\big)$ is weighed by the traffic demand $d_i$ to give more importance on busy APs and the expectation $\mathbb{E}_\bbd[\cdot]$ is w.r.t. the distribution of the traffic demands $\bbd$. Note that the objective function can be any combination of $\{u_i\big(\bbd, \ccalN, \bbC\big)\}_{i=1}^N$ that is not limited to \eqref{eq:objectiveFunction}. We formulate the channel management problem in WLANs as \vspace{-2mm}
\begin{alignat}{3} \label{eq:problemDefinition}
	\mathbb{P}&:=  \min_{\bbC} \  \mathbb{E}_\bbd\Big[\frac{1}{N}\sum_{i=1}^N d_i u_i\big(\bbd, \ccalN, \bbC\big) \Big],             \\
	&  \st \quad \bbc_i \in \{0,1\}^M,~\forall~i=1,\ldots,N,  \nonumber
\end{alignat}
where the channel selections $\bbC$ are controlled by the policy $\pi$ based on the traffic demands $\bbd$. While some theoretical models exist for the interference function $I_{i,j,\ell}(\cdot)$ in $u_i(\cdot)$ [cf. \eqref{eq:channelUtilization}] and the distribution of the traffic demands $\bbd$ in $\mathbb{E}_\bbd[\cdot]$, we stress that problem \eqref{eq:problemDefinition} is given without any explicit model because the latter may not be accurate for real-world systems leading to inevitable degradation of model-based approaches. While this is a centralized problem that integrates the global network information in the problem formulation, we aim to find a solution that allows for a decentralized implementation, i.e., the control policy can be shared over all APs and each AP can deploy it to allocate channels locally with only neighborhood information. These facts result in three main challenges for solving problem \eqref{eq:problemDefinition}:

\begin{enumerate}
	
	\item The problem requires a solution that can be implemented in a decentralized manner, where each AP communicates with its neighbors, aggregates the neighborhood information, and makes its own decisions.
	
	\item The problem requires a model-free solution that does not rely on theoretical system models, which precludes the application of model-based heuristic methods.
	
	\item The problem requires continuous relaxation of discrete actions to compute gradients for conventional learning methods, which leads to performance degradation. %degrades the raises the training difficulty of conventional learning methods.  
	
\end{enumerate}

In the next section, we propose a learning-based approach that allows decentralized channel allocation and solves the problem in a model-free manner.

%%%%%%%%%%%%%%%%%%%%%%%%%%%%%%%%%%%%
%%%%%%%%%%%%%%%%%%%%%%%%%%%%%%%%%%%%
%%%%  S  E  C  T  I  O  N     %%%%%%%%%%%%%%%%%%
%%%%%%%%%%%%%%%%%%%%%%%%%%%%%%%%%%%%
%%%%%%%%%%%%%%%%%%%%%%%%%%%%%%%%%%%%

\section{Methodology}

With the aforementioned challenges, we pursue to solve problem \eqref{eq:problemDefinition} with a learning procedure that combines \emph{graph neural networks (GNNs)} and the \emph{policy gradient} method \cite{policygradient}. The former allows for a decentralized implementation, while the latter yields a model-free solution. Specifically, we start by parameterizing the policy $\pi(\bbd)$ with GNNs and re-formulate problem \eqref{eq:problemDefinition} as a statistical learning problem. We proceed to solve this learning problem with the policy gradient method, which needs only system observations but not system models. 

\subsection{Graph Neural Networks (GNNs)}\label{sec_gnn}

A WLAN can be modeled by a graph $\ccalG = (\ccalV, \ccalE)$ with the node set $\ccalV$ representing APs and the edge set $\ccalE$ representing communication links. The graph structure is captured by a matrix $\bbS \in \mathbb{R}^{N \times N}$, referred to as the \emph{graph shift operator}, with the $(i,j)$th entry non-zero $[\bbS]_{ij} \ne 0$ if and only if there exists an edge between node $n_i$ and $n_j$, i.e., $(n_i,n_j)\in \ccalE$. The system states are modeled by a graph signal $\bbx \in \mathbb{R}^N$ with the $i$th entry $[\bbx]_i$ representing the state of node $n_i$, such as the traffic demands of APs $\bbd$. The graph signal exhibits an irregular structure inherent in the graph topology and the coupling between them can be exploited for feature extraction.

\smallskip
\noindent \textbf{Graph convolutional filters (GCFs).} The GCF is one of the key tools in graph signal processing, which leverages the structural information to model representations from graph signals \cite{gavili2017shift}. Specifically, the GCF is a polynomial of the shift operator matrix $\bbS$ over the graph signal $\bbx$ as \vspace{-1mm}
\begin{align}\label{eq:graphFilter}
	\bbH(\bbS)\bbx = \sum_{k=0}^K h_k \bbS^k \bbx
\end{align}
with $\{h_k\}_{k=0}^K$ the filter coefficients. The shift operation $\bbS \bbx$ assigns to node $n_i$ the aggregated information $[\bbS\bbx]_i$ from its neighboring nodes and $\bbS^k \bbx$ represents information exchanges up to a neighborhood of radius $k$ -- see Fig. 2. %In this context, 
The GCF is a shift-and-sum operator that generalizes the convolution from the time/image domain to the graph domain and generates features from a more complete picture of the graph.

\smallskip
\noindent \textbf{GNNs.} The GNN is a layered architecture, where each layer comprises a bank of GCFs and a pointwise nonlinearity \cite{henaff2015deep, defferrard2016convolutional, gao2021stochastic}. At layer $\ell = 1,\ldots, L$, there are $F_{\ell-1}$ input signals $\{\bbx_{\ell-1}^g\}_{g=1}^{F_{\ell-1}}$ generated at layer $\ell-1$. The latter are processed by a bank of GCFs $\{\bbH_{\ell}^{fg}(\bbS)\}_{fg}$, aggregated over the input index $g$, and passed through a pointwise nonlinearity $\sigma(\cdot):\mathbb{R} \to \mathbb{R}$ to generate the output signals of layer $\ell$ as \vspace{-1mm}
\begin{align}\label{eq:GNN}
	\bbx_{\ell}^f = \sigma\left(\sum_{g=1}^{F_{\ell-1}} \bbH_{\ell}^{fg}(\bbS) \bbx_{\ell-1}^g \right),~\forall~f=1,\ldots,F_\ell,
\end{align}
%to produce intermediate features as
%\begin{align}
%	\bbu_{\ell}^{fg} = \bbH_{\ell}^{fg}(\bbS) \bbx_{\ell-1}^g
%\end{align}
%for $f=1,...,F_\ell$ and $g=1,...,F_{\ell-1}$. These intermediate features are aggregated over the input index $g$ and passed through a pointwise nonlinearity $\sigma(\cdot):\mathbb{R} \to \mathbb{R}$ to generate the outputs of layer $\ell$ as
%\begin{align}\label{eq:GNN}
%	\bbx_{\ell}^f = \sigma\left(\sum_{g=1}^{F_{\ell-1}} \bbu_{\ell}^{fg}\right),~\forall~f=1,\ldots,F_\ell,
%\end{align}
where common examples of $\sigma(\cdot)$ include the ReLU, absolute value, etc. Without loss of generality, we assume a single input $\bbx_0^1 = \bbx$ and a single output $\bbx_L^1$. The overall GNN can be represented as a nonlinear mapping $\bbPhi(\bbx,\bbS,\ccalH)$, where the inputs are the graph signal $\bbx$, the shift operator $\bbS$, and the output is the task-relevant feature. The architecture parameters $\ccalH$ are the filter coefficients of GCFs throughout $L$ layers.

\smallskip
\noindent \textbf{Statistical learning problem.} By parameterizing the policy $\pi(\bbd)$ with the GNN $\bbPhi(\bbd, \bbS, \ccalH)$ of parameters $\ccalH$, we can re-formulate problem \eqref{eq:problemDefinition} as a statistical learning problem
\begin{alignat}{3} \label{eq:learningProblem}
	\mathbb{P}_\ccalH &:= \min_{\ccalH} \ \ccalL(\ccalH) =  \min_{\ccalH} \  \mathbb{E}_\bbd\Big[\frac{1}{N}\sum_{i=1}^N d_i u_i(\bbd, \bbS, \ccalH) \Big],             \\
	&  \st \quad [\bbPhi(\bbd, \bbS, \ccalH)]_i \in \{0,1\}^M,~\forall~i=1,\ldots,N.  \nonumber
\end{alignat}
The goal is to learn the optimal parameters $\ccalH^*$ that minimize the objective function. It is challenging to solve this problem by conventional unsupervised learning methods because they (i) rely on explicit theoretical models of the interference function $I_{i,j,\ell}(\cdot)$ and the traffic demand distribution $\mathbb{E}_\bbd[\cdot]$; (ii) require well-designed continuous relaxation for the discrete action space, to make gradient computations possible. Motivated by these challenges, we propose to leverage the model-free policy gradient method, which is widely used in reinforcement learning, to conduct the training procedure.

\subsection{Policy Gradient}\label{sec:policyGradient}

The training procedure optimizes the GNN parameters $\ccalH$ with gradient descent. At each iteration $t$, we update the current parameters $\ccalH^{(t)}$ as \vspace{-1mm} 
\begin{align} \label{eq:update}
	&\ccalH^{(t+1)} = \ccalH^{(t)} - \alpha^{(t)} \nabla_{\ccalH} \mathcal{L}(\ccalH^{(t)}),
\end{align}
where $\alpha^{(t)} > 0$ is the step-size. The preceding update requires the system model and the continuous relaxation to compute the gradient $\nabla_{\ccalH} \mathcal{L}(\ccalH^{(t)})$, where the former is not assumed available in problem \eqref{eq:learningProblem} and the latter leads to performance degradation. We overcome these issues by leveraging the policy gradient method. It exploits the likelihood ratio property to provide a stochastic approximation for the gradient of policy functions that take the form of $\mathbb{E}[F(\bbd,\bbPhi(\bbx,\bbS,\ccalH))]$, where $F(\cdot)$ is a general unknown function \cite{sutton1999policy}.

Specifically, the policy gradient method considers the GNN outputs $\bbPhi(\bbd,\bbS,\ccalH)$ as samples drawn from a probability distribution $p_{\bbd,\bbS,\ccalH}(\bbPhi)$ specified by $\bbd$, $\bbS$ and $\ccalH$, which has a delta density function $f_{\bbd,\bbS,\ccalH}(\bbPhi) = \delta(\bbPhi - \bbPhi(\bbd,\bbS,\ccalH))$. This allows to compute the Jacobian of the policy function as \vspace{-1mm}
\begin{align} \label{eq:policyGradient}
		\nabla_{\ccalH} \mathbb{E} [F(\bbd,\!\bbPhi(\bbd,\!\bbS,\!\ccalH)\!)] \!=\! \mathbb{E}[F(\bbd,\! \bbPhi) \nabla_{\ccalH}\! \log\! f_{\bbd,\bbS,\ccalH}(\bbPhi\!)].
\end{align}
It translates the computation of $\nabla_{\ccalH} \mathbb{E} [F(\bbd,\bbPhi(\bbd,\!\bbS,\!\ccalH))]$ to the multiplication of $F(\bbd, \bbPhi)$ and $\nabla_{\ccalH}\! \log\! f_{\bbd,\bbS,\ccalH}(\bbPhi)$. However, computing the gradient of the delta density function $\nabla_{\ccalH}\! \log\! f_{\bbd,\bbS,\ccalH}(\bbPhi)$ requires the system model $F(\cdot)$ as well. We resolve this issue by approximating the delta density function with another density function centered around $\bbPhi(\bbd,\bbS,\ccalH)$. Examples include the Gaussian distribution, binomial distribution and categorical distribution. Then, we can estimate the gradient of the loss function $\nabla_{\ccalH} \mathcal{L}(\ccalH^{(t)})$ as \vspace{-1mm}
\begin{align}\label{eq:policyGradient2}
	&\nabla_{\ccalH} \mathcal{L}(\ccalH^{(t)}) \\
	&= \mathbb{E}_\bbd\Big[\frac{1}{N}\sum_{i=1}^N d_i u_i(\bbd, \bbS, \ccalH^{(t)}) \nabla_{\ccalH}\! \log\! f_{\bbd,\bbS,\ccalH^{(t)}}(\bbPhi)\Big], \nonumber
\end{align}
where the expectation $\mathbb{E}_\bbd[\cdot]$ is approximated by observing $\ccalT$ samples of the traffic demands $\{\bbd_\tau\}_{\tau=1}^\ccalT$ and taking the average. We remark that \eqref{eq:policyGradient2} is model-free, i.e., it does not require any theoretical model of the interference function $u_i(\cdot)$ but only function values $u_i(\bbd_\tau, \bbS, \ccalH^{(t)})$ evaluated at the sampled traffic demands $\bbd_\tau$, %and actions $\bbPhi_\tau$, 
which can be observed from the system directly. Moreover, it deals with the discrete action space by sampling actions from policy distributions, e.g., binomial distributions and categorical distributions, which avoids the performance degradation induced by the continuous relaxation in conventional unsupervised learning methods.

\subsection{Decentralized Implementation}\label{sec:decentralizedImplementation}

The trained GNN admits a decentralized implementation, i.e., each node can compute outputs and make decisions locally with neighborhood information. To see this, let $[\bbS]_{ij}$ be the $(i,j)$th entry of $\bbS$ and $[\bbx]_i$ the $i$th entry of $\bbx$. The graph sparsity allows to compute the shift operation $\bbS\bbx$ as
\begin{align}
	[\bbS\bbx]_i = \sum_{j=1}^N [\bbS]_{ij} [\bbx]_j = \sum_{j: \ \ccalN(i,j)=1} [\bbS]_{ij} [\bbx]_j
\end{align}
for $i=1,\ldots,N$. That is, the entry $[\bbS \bbx]_i$ at node $n_i$ can be computed with the entries $\{[\bbx]_j\}_{j}$ at nodes $\{n_j\}_j$ that are connected to node $n_i$. This indicates that the entries of $K$ shifted signals $\{[\bbS^k \bbx]_i\}_{k=0}^K$ can be computed at node $n_i$ by exchanging information with its neighboring nodes as well. Since the aggregation of the filter coefficients $\{h_k\}_{k=0}^K$ is a linear combination of signal values at individual node, which dose not involve any feature fusion among nodes, the entry of the filter output $[\bbH(\bbS)\bbx]_i$ at node $n_i$ can be computed with local neighborhood information; hence, GCFs allow for a decentralized implementation. %\cite{segarra2017optimal}.

GNNs inherit the decentralized implementation from GCFs because the nonlinearity $\sigma(\cdot)$ is pointwise and local \cite{li2020graph, gao2022wide, hansen2022power}. This analysis shows that while GCFs and GNNs are expressed in compact forms \eqref{eq:graphFilter} and \eqref{eq:GNN}, they can be implemented in a decentralized manner, i.e., each node does not need to know the entire graph topology, but only have the communication capacity to exchange information with neighbors and the computation capacity to aggregate the neighborhood information.

\subsection{Permutation Equivariance} \label{sec:permutationEquivariance}

GNNs account for the graph structure to extract features and thus, are equivariant to topology permutations, i.e., the node reordering. This is a key property that allows GNNs to exploit internal symmetries of graph signals, which demonstrates the advantage of GNNs to process networked data and justifies the effectiveness of the proposed approach. Specifically, define the permutation matrix as $\bbP \in \{ 0,1 \}^{N \times N}$ with $\bbP \bm{1} = \bm{1}$ and $\bbP^\top \bm{1} = \bm{1}$. %as
%\begin{equation} \label{eq:permutationDefinition}
%	\begin{split}
%		\bbP \in \{ 0,1 \}^{N \times N}: \bbP \bm{1} = \bm{1}, \quad \bbP^\top \bm{1} = \bm{1}.
%	\end{split}
%\end{equation}
The permuted matrix $\bbP \bbS \bbP^\top$ reorders the columns and rows of $\bbS$, while the permuted vector $\bbP \bbx$ reorders the entries of $\bbx$. According to \cite{gama2020stability, gao2021stability}, the following theorem formally characterizes the permutation equivariance of GNNs.
\begin{theorem}[Proposition 1 \cite{gama2020stability}]\label{thm:permutationEquivariance}
	Consider the GNN $\bbPhi(\bbx,\bbS,\ccalH)$ with the graph signal $\bbx$ and the underlying graph $\bbS$. %and the architecture parameters $\ccalH$. 
	Let $\bbP$ be a permutation matrix. Then, it holds that 
	%signal $\bbx$, graph matrix $\bbS$ and parameters $\bbtheta$. For any permutation $\bbPi$, it holds that
	\begin{equation} \label{eq:permutationEquivariance}
		\begin{split}
			\bbPhi(\bbP \bbx,\bbP\bbS \bbP^\top ,\ccalH) = \bbP \bbPhi(\bbx,\bbS,\ccalH).
		\end{split}
	\end{equation}
\end{theorem}
Theorem \ref{thm:permutationEquivariance} states that GNNs applied on permuted graphs $\bbP \bbS \bbP^\top$ and permuted signals $\bbP \bbx$ generate equally permuted outputs. This result indicates that GNNs can harness the same information regardless of what permuted versions of graphs and signals are processed. The latter acts as a data augmentation technique that facilitates the training procedure, which will be validated in Section \ref{sec_numerical_results}. In the context of channel allocation, this property yields the implementation of GNN-based policies independent of the order of APs, i.e., they can be implemented directly on arbitrarily ordered APs. This follows the intuitive requirement of decentralized channel management solutions in WLANs because APs will not be labeled with a fixed order in practice and the explicit order is not the prior information available during implementation.

 %%%%%%%%%%%%%%%%%%%%%%%%%%%%%%%%%%%%%%%%%%%%%%%%%%%%%%%%%%%%%%%%%%%%
%%%   S   E   C   T   I   O   N   %%%%%%%%%%%%%%%%%%%%%%%%%%%%%%%%%%
%%%%%%%%%%%%%%%%%%%%%%%%%%%%%%%%%%%%%%%%%%%%%%%%%%%%%%%%%%%%%%%%%%%%

\section{Results}\label{sec_numerical_results}

\begin{figure}
	\centering
	\includegraphics[width=0.3\textwidth, trim=0 0 0 0]{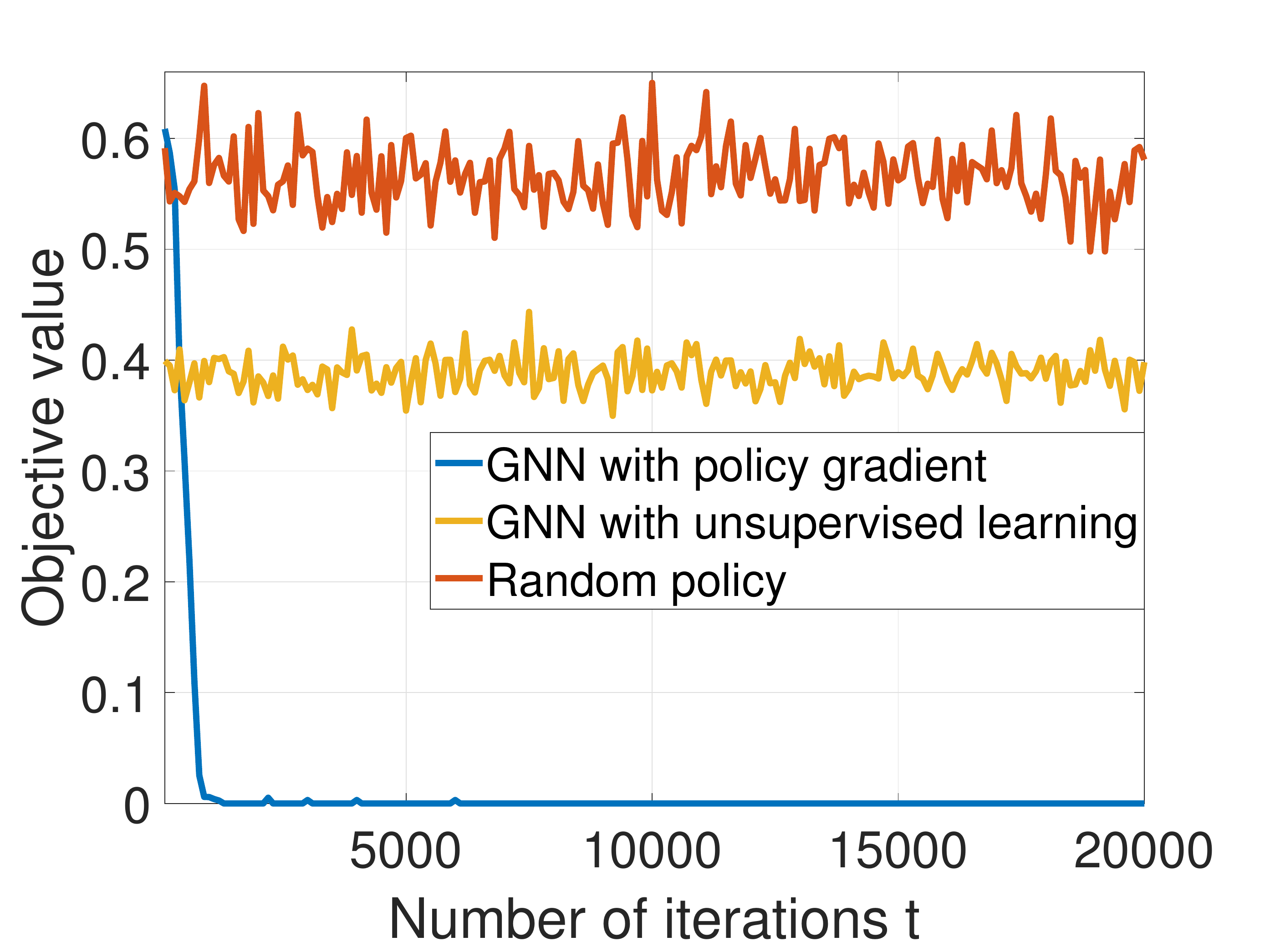}
	\caption{Performance of GNN with policy gradient compared to GNN with conventional unsupervised learning and the random channel selection policy.}
	\label{fig:example}\vspace{-6mm}
\end{figure}
 
In this section, we present empirical results for performance evaluation. While the proposed approach is model-free, we make system observations (i.e., observations of the traffic demands and the interference) from given models. However, we do not assume any knowledge of these models during training/testing but only generate samples %from them 
for observation.  

\smallskip
\noindent \textbf{Setup.} The network consists of $N$ APs and the topology is generated randomly with the edge probability $q$, i.e., AP $n_i$ is connected to AP $n_j$ with probability $q$ for $i,j=1,\ldots,N$. Traffic demands are sampled from distribution $\max(\mathbb{N}(0.8,0.4),0)$, where $\mathbb{N}(\mu,\rho)$ denotes the normal distribution with mean $\mu$ and standard deviation $\rho$, and the interference is derived from the model $I_{i,j,\ell} = \ccalN(i,j) c_{i\ell}c_{j\ell} d_j / M_j$. The GNN has $4$ layers, each containing $32, 64, 64, 32$ filters of order $3$ and the ReLU nonlinearity. The policy distribution is the categorical distribution and the batch size is $64$. Our results are averaged over $10$ randomly generated graphs.

\smallskip
\noindent \textbf{Small-scale scenario.} First, we consider a small-scale scenario with $10$ APs, $4$ channels and $q=0.25$ edge probability, where zero interference can be theoretically achieved by allocating different channels to adjacent APs and the same channels to nonadjacent APs. We compare the proposed approach with two baselines: (i) GNN with conventional unsupervised learning \cite{kaushik2021deep}, and (ii) uniformly random channel selection. The former relaxes discrete actions as continuous variables for training and conducts discrete quantification for testing, which is a model-based solution that depends on closed-form system models to compute gradients, while the latter is model-free. %considers all channel allocation plans and selects one of them randomly at each AP, which is a model-free solution. 
Fig. \ref{fig:example} shows the results. The proposed approach exhibits the optimal performance, i.e., zero interference among APs. The GNN with unsupervised learning performs worse despite the model-based property. This is because it approximates discrete actions via a continuous relaxation, and hence, suffers from performance degradation. Moreover, we see from the figure that it does not show learning behaviour. %from its curve 
This is because (i) the maximal operation in \eqref{eq:channelUtilization} makes the gradient propagation difficult even for continuous variables, and (ii) the discrete quantification during testing creates further difficulty. %interferes with the curve. 
%Both GNN methods outperform the random policy. 

\begin{figure}%
	\centering
	\begin{subfigure}{0.49\columnwidth}
		\includegraphics[width=1.0\linewidth, height = 0.775\linewidth]{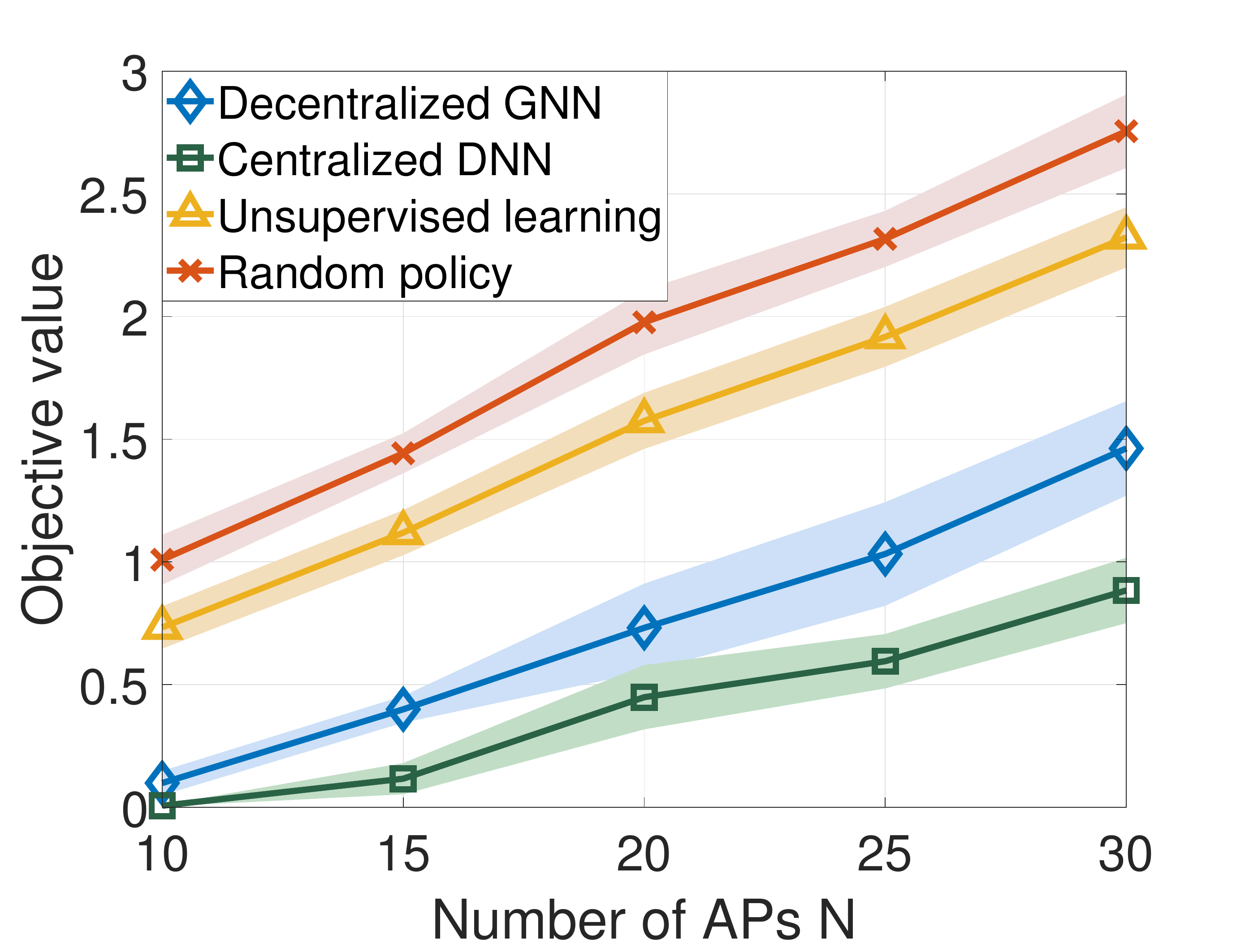}%
		\caption{}%
		\label{subfig:varyingNodes}%
	\end{subfigure}\hfill\hfill%
	\begin{subfigure}{0.49\columnwidth}
		\includegraphics[width=1.0\linewidth,height = 0.775\linewidth]{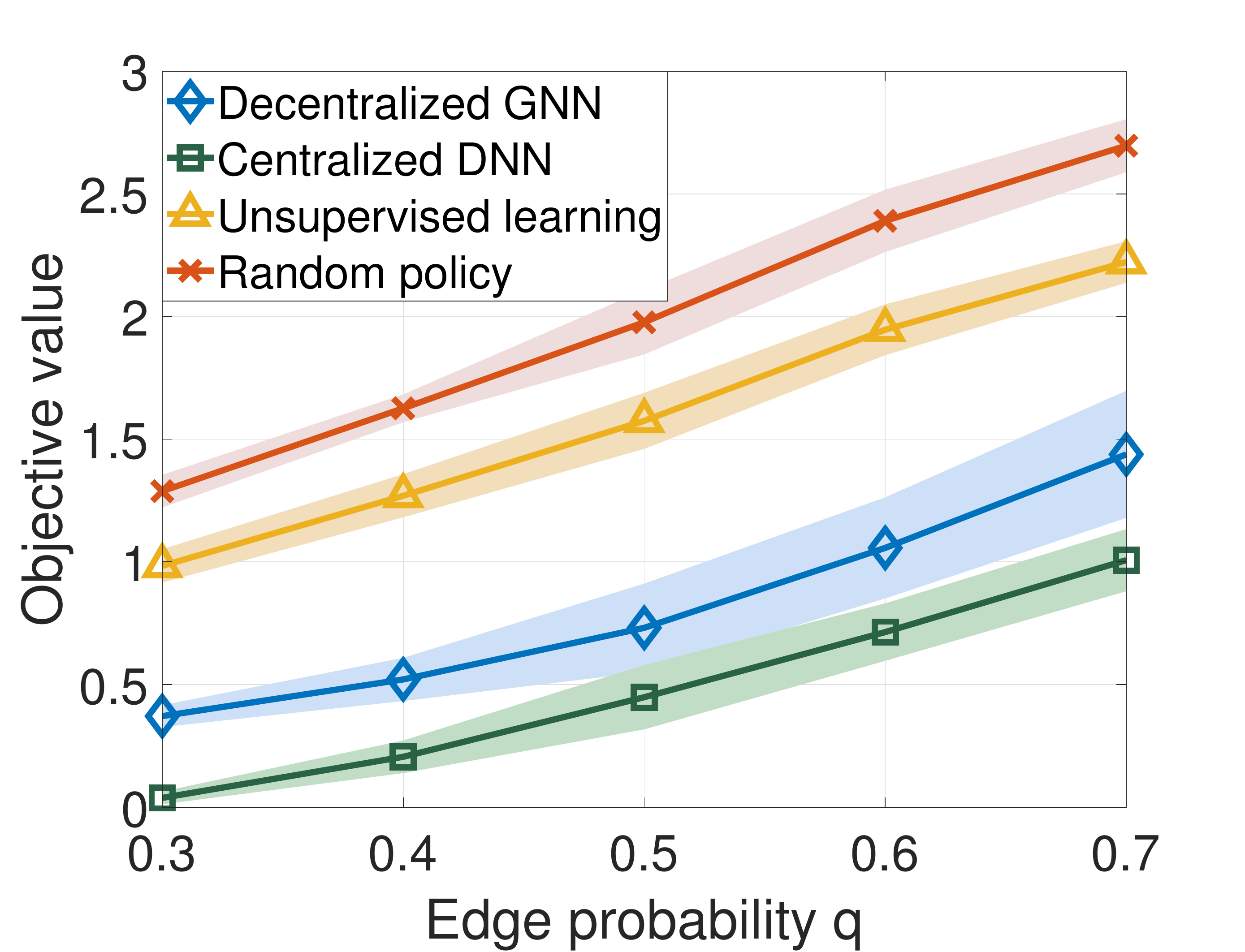}%
		\caption{}%
		\label{subfig:varyingEdges}%
	\end{subfigure}
	\caption{Performance of decentralized GNN compared with the baseline policies. The results are averaged over $10$ random networks and the shaded area shows the std. dev. (a) Different numbers of APs, i.e., different network sizes, with $q=0.5$ edge probability. (b) Different edge probabilities $q$, i.e., different network densities, with $20$ APs.}\label{fig:varyingNetworks}\vspace{-6mm}
\end{figure}

\smallskip
\noindent \textbf{Large-scale scenario.} Next, we evaluate the proposed approach on larger networks with various numbers of APs in Fig. \ref{subfig:varyingNodes} and various numbers of edges in Fig. \ref{subfig:varyingEdges}. The former expands the network size, while the latter increases the network density. We consider three baselines for performance comparison: (i) deep neural network (DNN), (ii) GNN with unsupervised learning, and (iii) random channel selection. The first follows \cite{HuaweiFrance} to parameterize the control policy with a DNN, which is a centralized solution that requires the global network information %the central controller for channel allocation 
during implementation, while the latter two are decentralized solutions. The centralized DNN performs best but its execution requires access to the states, i.e., the traffic demands, %global information 
of all the APs, which may not be practical for large-scale scenarios. The proposed approach achieves a performance close to the one achieved by the centralized DNN, but requires only local neighborhood information for implementation, yielding a satisfactory trade-off between performance and scalability. Moreover, the proposed approach generalizes well for large and dense networks, in which decentralization plays an important role.

%\smallskip
%\noindent \textbf{Training efficiency.} Next, we compare the training efficiency between the proposed approach and the centralized DNN. In Fig. \ref{subfig:trainingProcedure}, we see that the proposed approach converges much faster than the centralized DNN and shows better training efficiency. The result corroborates the fact that the GNN is permutation equivariant and the latter acts as a data augmentation technique to facilitate the training procedure, as indicated in Theorem \ref{thm:permutationEquivariance}.

\begin{figure}%
	\centering
	\begin{subfigure}{0.49\columnwidth}
		\includegraphics[width=1.0\linewidth, height = 0.775\linewidth]{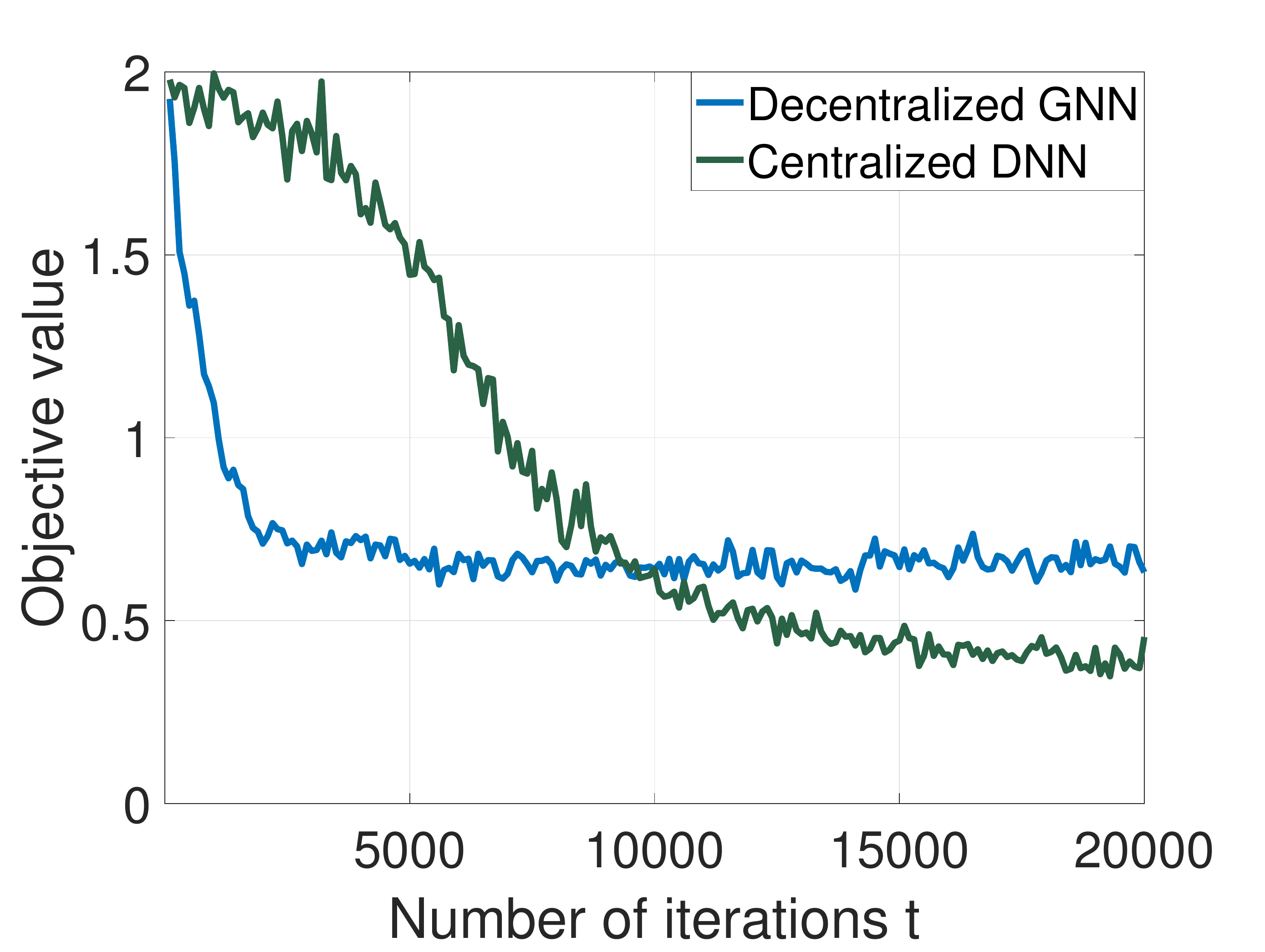}%
		\caption{}%
		\label{subfig:trainingProcedure}%
	\end{subfigure}\hfill\hfill%
	\begin{subfigure}{0.49\columnwidth}
		\includegraphics[width=1.0\linewidth,height = 0.775\linewidth]{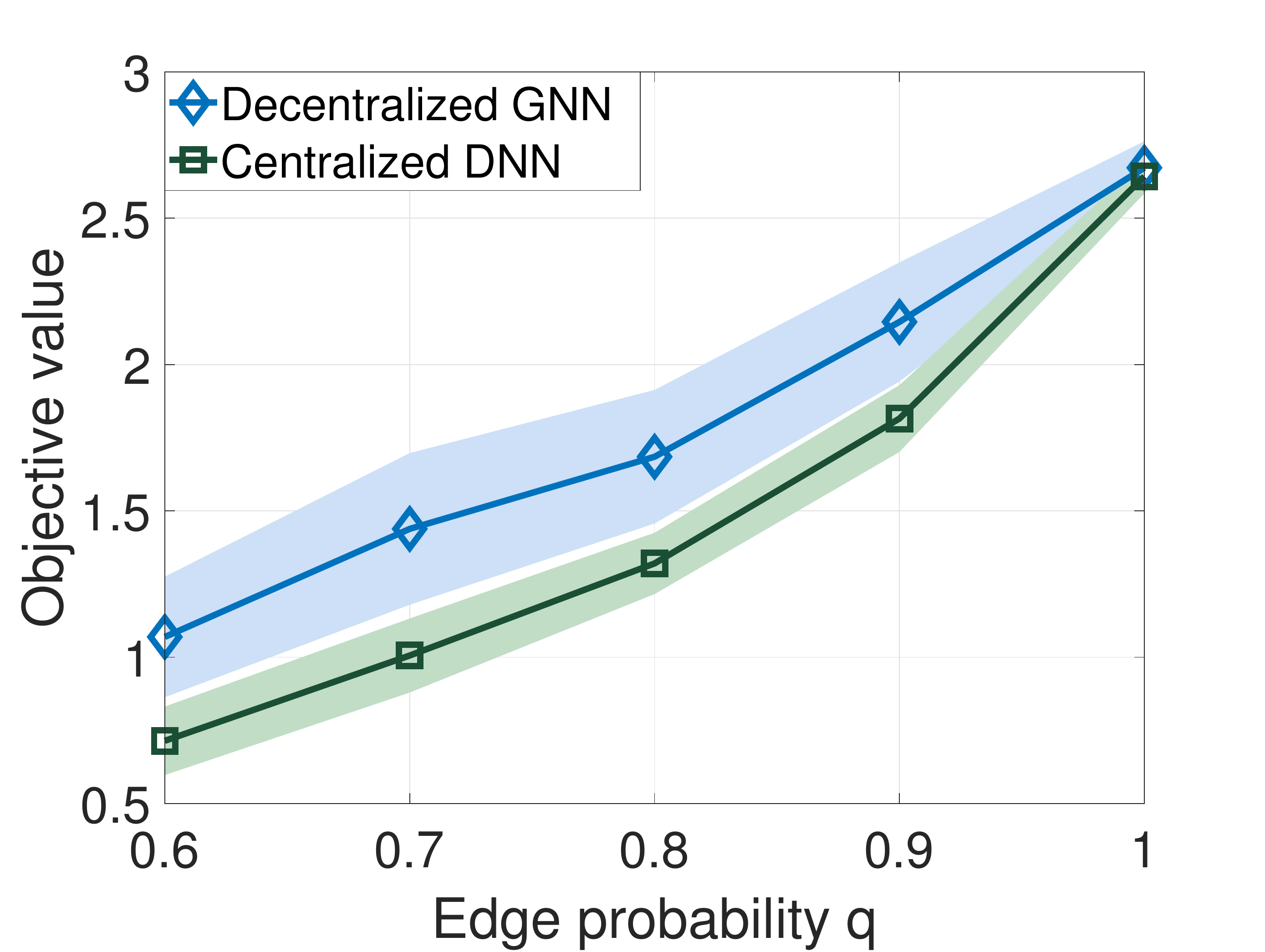}%
		\caption{}%
		\label{subfig:approach}%
	\end{subfigure}
	\caption{(a) Training performance of decentralized GNN and centralized DNN for a network with $20$ APs and $q=0.5$ edge probability. (b) Performance of decentralized GNN and centralized DNN for a network with $20$ APs and different graph connections, i.e., different edge probabilities $q$.}\label{fig:extraExperiment}\vspace{-4mm}
\end{figure}

\smallskip
\noindent \textbf{GNN v.s. DNN.} Next, we compare the proposed approach with the centralized DNN in terms of the training efficiency and the graph connection. 
%with training efficiency between the proposed approach and the centralized DNN. 
In Fig. \ref{subfig:trainingProcedure}, we see that the proposed approach converges much faster than the centralized DNN and shows better training efficiency. The result corroborates the fact that the GNN is permutation equivariant and the latter acts as a data augmentation technique to facilitate the training procedure, as indicated in Theorem \ref{thm:permutationEquivariance}. In Fig. \ref{subfig:approach}, we see that the performance degradation of the decentralized GNN w.r.t. the centralized DNN 
%induced by the decentralized implementation 
first increases with the graph connectivity, i.e., from $0.355$ at $q = 0.6$ to $0.433$ at $q = 0.7$. This is because the interference becomes more severe and the global information becomes more important for channel allocation. The performance degradation then decreases from $q = 0.7$ to $q = 1$ and converges to almost zero for fully-connected graphs. We attribute this behaviour to the fact that the neighborhood information becomes richer as the graph connectivity increases, and the proposed approach performs comparably well with the centralized DNN.

\begin{figure}
	\centering
	\includegraphics[width=0.6\linewidth, height = 0.45\linewidth, trim=0 0 0 0]{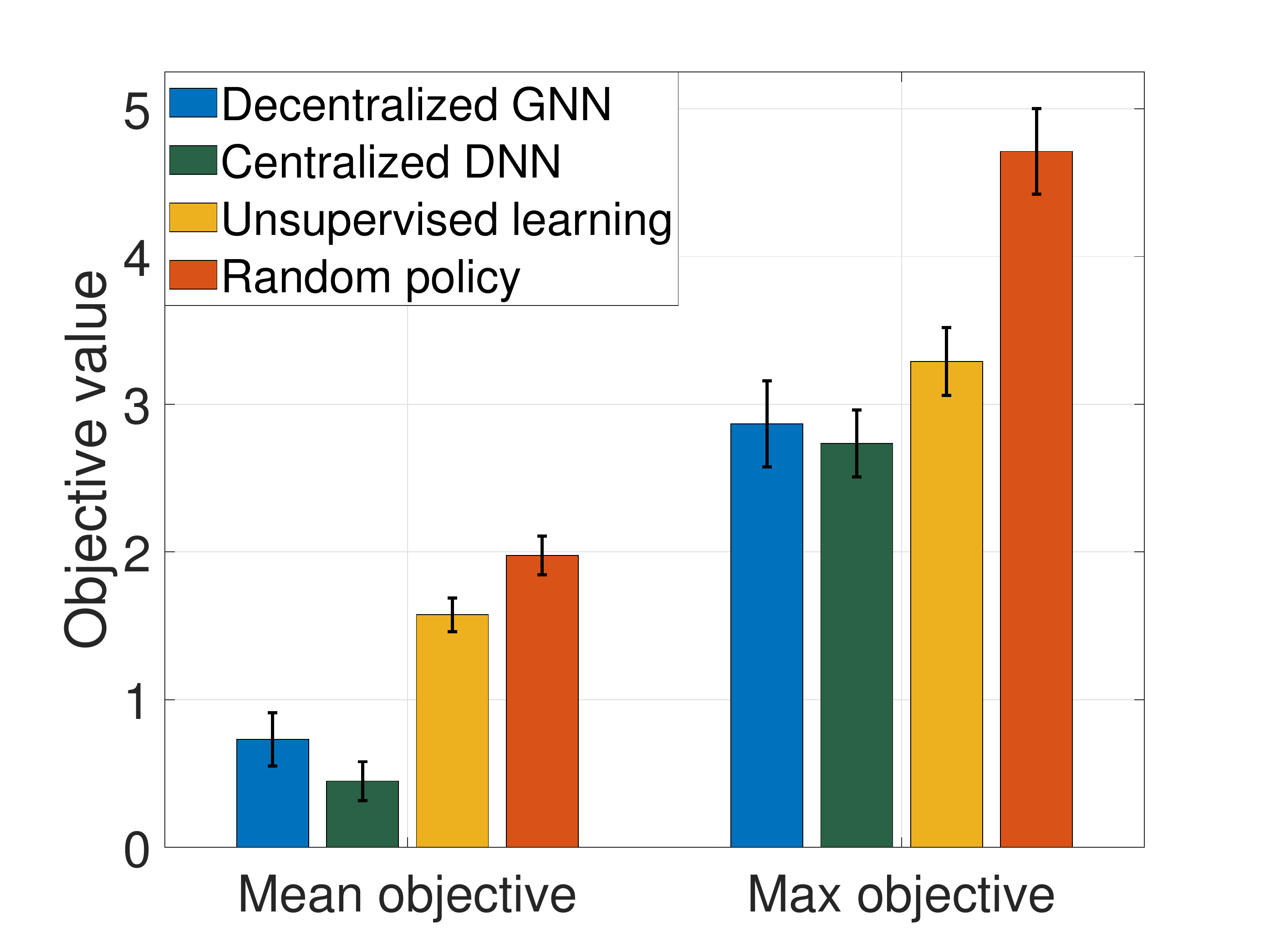}
	\caption{Performance of decentralized GNN compared with the baseline policies w.r.t. the mean and max objectives. The network has $20$ APs and $q=0.5$ edge probability, and the results are averaged over $10$ random networks.%the GNN and the DNN for the network with $20$ APs and different graph connections, i.e., different edge probabilities (high values of $q$).
	}
	\label{fig:meanMax}\vspace{-6mm}
\end{figure}

\smallskip
\noindent \textbf{Optimization criteria.} Lastly, we show the adaptivity of the proposed approach by testing a different objective function. We consider the objective function as the expected maximum of the maximal channel utilization over APs, i.e., the maximal channel utilization on the worst AP. The results are shown in Fig. \ref{fig:meanMax}. The proposed approach performs close to the centralized solution on both objective functions, indicating its capacity in handling different optimization criteria specified by different application requirements.

 %%%%%%%%%%%%%%%%%%%%%%%%%%%%%%%%%%%%%%%%%%%%%%%%%%%%%%%%%%%%%%%%%%%%
%%%   S   E   C   T   I   O   N   %%%%%%%%%%%%%%%%%%%%%%%%%%%%%%%%%%
%%%%%%%%%%%%%%%%%%%%%%%%%%%%%%%%%%%%%%%%%%%%%%%%%%%%%%%%%%%%%%%%%%%%

\section{Conclusion} \label{sec_con}
 
In this paper, we investigated the channel allocation problem in WLANs. We formulated a stochastic optimization problem, parameterized the allocation policy with GNNs, and transformed the problem to an unsupervised learning problem. We conducted training with the policy gradient method, which requires only system observations instead of system models, and can be adapted to application scenarios where system models are inaccurate or unknown. The proposed approach allows for a decentralized implementation by leveraging the distributed nature of GNNs, where each AP selects channels locally with only neighborhood information. This yields a scalable solution for large networks with no need of a central access controller during execution. We further show the permutation equivariance of the proposed approach, which acts as a data augmentation technique for training efficiency and makes the algorithm implementation independent of the AP reordering. Numerical experiments demonstrated the superior performance of the proposed approach, i.e., the interference among APs can be reduced in a decentralized fashion.

%\urlstyle{same}
\bibliographystyle{IEEEtran}
\bibliography{FSO_learning,wireless_learning}

\end{document}

%% file: my_sections.tex
\usepackage{needspace}